\documentclass[prd,aps,preprintnumbers,twocolumn,floatfix]{revtex4}
\usepackage{epsfig}
\usepackage{amsmath}

\def\K{{\mathcal K}}
\def\G{{\mathcal G}}
\def\M{{\mathcal M}}

\def\as{\ensuremath{\alpha_{s}}}
\def\a0{\alpha_0}

\def\vep{\varepsilon}
\def\f{{\rm f}}

\def\bea {\begin{eqnarray}}
\def\eea {\end{eqnarray}}

\def\be {\begin{equation}}
\def\ee {\end{equation}}

\begin{document}

\preprint{YITP-SB-06-25}
\preprint{SLAC--PUB--11907} 

\renewcommand{\thefigure}{\arabic{figure}}

\title{The Two-loop Anomalous Dimension Matrix for Soft Gluon Exchange}

\author{S.\ Mert Aybat$^1$, Lance J. Dixon$^2$, George Sterman$^1$}

\affiliation{${}^1$C.N.\ Yang Institute for Theoretical Physics, 
Stony Brook University, Stony Brook, New York 11794--3840, USA,}
\affiliation{$^{2}$ Stanford Linear Accelerator Center, Stanford University,
  Stanford, CA 94309, USA}
\date{\today}

\begin{abstract}
The resummation of soft gluon exchange for QCD hard scattering
requires a matrix of  anomalous dimensions.
We compute this matrix directly for arbitrary $2\to n$
massless processes for the first time at two loops.
Using color generator notation, we show
that it is proportional to the one-loop matrix.  
This result reproduces all pole terms in dimensional
regularization of the explicit calculations of massless
$2\to 2$ amplitudes in the literature, and it predicts
all poles at next-to-next-to-leading order in any $2\to n$ process 
that has been computed at next-to-leading order.
The proportionality of the one- and two-loop matrices 
makes possible the resummation in closed form of the
next-to-next-to-leading logarithms and poles
in dimensional regularization for the $2 \to n$ processes.
\end{abstract}

\maketitle

The calculation of high-energy cross sections in perturbative 
quantum chromodynamics (QCD) for hadronic collisions involves
the factorization of long- and short-distance effects.
Sensitivity to long-distance dynamics is enhanced 
by powers of logarithms whenever there is an incomplete cancellation
of parton emission and virtual corrections. 
In such situations, it is useful to organize, or resum, these
corrections to all orders in perturbation theory.
Correspondingly, in partonic scattering
or production amplitudes, it is necessary to organize poles in $\vep$
that arise in dimensional regularization (with $D=4-2\vep$).   
The resummation of these poles and related logarithmic enhancements
is well-understood for inclusive reactions mediated by electroweak
interactions, such as the Sudakov form factor~\cite{Sudrefs,magnea90}
and in Drell-Yan processes~\cite{kidonakisrv}.  With
recent advances in the computation of splitting functions~\cite{MVVsplit},
many such corrections can be resummed explicitly to 
next-to-next-to-leading level.  Their structure at
arbitrary level is known to be determined by a handful of
anomalous dimensions.  

The situation for QCD hard scattering processes containing four
or more partons --- critical to understanding many types of 
backgrounds to new physics at hadron colliders~\cite{LHCQCD} ---
is more complex.  Resummation beyond leading 
logarithms or poles requires a matrix of additional 
anomalous dimensions~\cite{matrixdim,KOS,KOSjet,BSZ}.  These matrices
are found in turn from the renormalization of the vacuum
matrix elements of products of Wilson lines, one
for each external parton in the underlying process~\cite{KOS}.
In this paper, we investigate the structure of the
two-loop anomalous dimension matrix.  We will find that,
remarkably, for every hard-scattering process involving
only massless partons, this matrix is proportional to the 
one-loop matrix.  We will concentrate below on the role 
that the matrix plays in partonic amplitudes.  
The full calculation of the two-loop matrix will be given
elsewhere~\cite{aybat06}.  In this paper, we provide the simple
calculation that is at the heart of the main result.  We will show
that certain color correlations due to two-loop diagrams that couple three
Wilson lines vanish identically.  We will also provide an
explicit expression in terms of color generators~\cite{catani96,catani98} 
for all single-pole terms in massless $2\to n$ amplitudes.

We consider a general process involving the scattering
of massless partons, which we denote by ``f":
\bea
\label{partproc}
\f : \quad f_1(p_1,r_1) + f_2(p_2,r_2) &\ &\nonumber\\
&\ & \hspace{-40mm}
\to f_3(p_3,r_3)  + \dots + f_{n+2}(p_{n+2},r_{n+2})\,.
\eea
The $f_i$ are the flavors of the participating partons, 
which carry momenta $\{ p_i \}$ and color $\{r_i\}$. Adopting the 
color-state notation of Ref.~\cite{catani98}, we represent 
the amplitude for this process as $|\M_\f \rangle$.

It is convenient to express these amplitudes as vectors
with $C$ elements in the space of color tensors, for some
choice of basis tensors $\{ \left(c_I\right)_{\{r_i\}} \}$~\cite{KOS,catani98,TYS},
\begin{eqnarray}
\label{amp}
\left|\, 
\M_\f \left(\beta_j,\frac{Q^2}{\mu^2},\as(\mu),\vep \right) 
\right\rangle
&\equiv& \nonumber\\
&\ & \hspace{-40mm}
\sum_{L=1}^C \M_{\f,L}\left(\beta_j,\frac{Q^2}{\mu^2},\as(\mu),\vep \right)
\, \left(c_L\right)_{\{r_i\}}
\nonumber\\
\, .
\end{eqnarray}
We will analyze these amplitudes at fixed momenta $p_i$ 
for the participating partons, which we represent as
$p_i = Q\beta_i \, , \quad \beta_i^2=0\, ,$
where the $\beta_i$ are four-velocities, and where
$Q$ is an overall momentum scale.  

In dimensional regularization, on-shell amplitudes may be
factorized into jet, soft and hard functions that describe the
dynamics of partons collinear with the external  lines, soft exchanges
between those partons, and the short-distance scattering process,
respectively.  This factorization follows from the general
space-time structure of long-distance contributions
to elastic processes~\cite{Akhoury,matrixdim}.  
The general form of the factorized amplitude, for
equal factorization and renormalization scales $\mu$, 
is~\cite{TYS} 
\begin{eqnarray}
\label{facamp}
\left |\, {\cal M}_\f
  \left(\beta_i,\frac{Q^2}{\mu^2},\as(\mu),\vep \right)\right \rangle
&=&\prod_{i=1}^{n+2} J^{[i]}\left(\as(\mu),\vep
\right) 
\nonumber\\
&\ & \hspace{-45mm}
\times\, {\bf S}_\f\left( \beta_i,\frac{Q^2}{\mu^2},\as(\mu),
                     \vep \right) 
\left| \, H_\f
  \left(\beta_i,\frac{Q^2}{\mu^2},\as(\mu)\right) \right \rangle
\,,~~~
\end{eqnarray}
where $J^{[i]}$ is the jet function for external parton $i$, 
${\bf S}_\f$ is the soft function, and $H_\f$ is the 
hard (short-distance) function.

The jet function for parton $i$ can be expressed to all orders in terms
of three anomalous dimensions, ${\cal K}^{[i]}$, ${\cal G}^{[i]}$
and $\gamma_K^{[i]}$, of which the first is determined
order-by-order from the third.  The general form
of the jet function, and its expansion to second order is given by
(expanding any function as 
$f(\as)=\sum_n (\as/\pi)^n\, f^{(n)}$)~\cite{magnea90},
\begin{widetext}
\begin{eqnarray}
\ln \label{solutionev} 
J^{[i]}\left(\as(\mu),\vep\right)
&=& 
~\frac{1}{2} \int_{0}^{\mu}\frac{d\xi}{\xi}
\Biggl[
\K^{[i]}(\as(\mu),\vep)
+\,\G^{[i]}
\left(-1,\bar\as\left(\xi,\vep\right),\vep\right)
+\, \int_{\xi}^{\mu} 
\frac{d\tilde{\mu}}{\tilde{\mu}}
\gamma^{[i]}_{K}\left(\bar\as\left(\tilde{\mu},\vep
  \right)\right)
\Biggr] 
\\
&\ & \hspace{-31mm}=  
  - \left( \frac{\as}{\pi} \right)
\left(\, \frac{1}{8\varepsilon^2}\, \gamma^{[i] (1)}_K + 
  \frac{1}{4\varepsilon} {\cal G}^{[i] (1)}(\varepsilon)\, \right)
 +  \left( \frac{\as}{\pi} \right)^2
\left[ \frac{\beta_0}{32}\,\frac{1}{\varepsilon^2}
 \, \left( \frac{3}{4\varepsilon}\gamma_K^{[i] (1)}  
     + {\cal G}^{[i] (1)}(\varepsilon)\, \right)
 - \frac{1}{8}\left( \, \frac{\gamma_K^{[i] (2)}}{4\varepsilon^2}
  + \frac{{\cal G}^{[i] (2)}(\varepsilon)}{\varepsilon}
 \right) \, \right] \, + \dots\,\,. \nonumber
\end{eqnarray}
In the expansion we use the $D$-dimensional running-coupling, 
evaluated at one-loop order,
\begin{eqnarray}
\label{asinD}
\bar\as\left(\tilde{\mu},\vep\right) &=&
\as(\mu)\left(\frac{\mu^2}{\tilde{\mu}^2} \right)^\vep
\sum_{n=0}^{\infty}\left[\frac{\beta_0}{4\pi\vep}
\left(\left(\frac{\mu^2}{\tilde{\mu}^2} \right)^\vep-1\right) \as(\mu)
\right]^n \,,
\end{eqnarray}
with the one-loop coefficient
$\beta_0 = 11 C_A/3 - 4 T_F n_F/3$.
The corresponding expression for the soft matrix is
\bea
\label{expoS}
{\bf S}_\f \left(\frac{\beta_i\cdot \beta_j}{u_0},\as(\mu),\vep \right)
 &=&
{\rm P}~{\rm exp}\left[
\, -\; \int_{0}^{\mu} \frac{d\tilde{\mu}}{\tilde{\mu}}
{\bf \Gamma}_{S_\f} \left({\beta_i\cdot\beta_j \over u_0} ,
  \bar\as\left(\tilde{\mu}, \vep\right)\right) \right]
\nonumber\\
&=&
 1+\frac{1}{2\vep}\left(\frac{\alpha_s}{\pi}\right){\bf \Gamma}_{S_\f}^{(1)}\,
+\frac{1}{8\vep^2}\left(\frac{\alpha_s}{\pi}\right)^2
\left({\bf \Gamma}_{S_\f}^{(1)}\right)^2
- \frac{\beta_0}{16\vep^2}
\, \left(\frac{\alpha_s}{\pi}\right)^2{\bf \Gamma}_{S_\f}^{(1)}\,
+\frac{1}{4\vep}\,
\left(\frac{\alpha_s}{\pi}\right)^2
{\bf \Gamma}_{S_\f}^{(2)} + \dots\, ,~~~
\end{eqnarray}
\end{widetext}
where $u_0 = \mu^2/Q^2$, so that $\beta_i\cdot\beta_j/u_0 = s_{ij}/\mu^2$.

Expanding ${\cal G}^{[i]}={\cal G}_0^{[i]}+\vep{\cal G}^{[i]\prime}+\dots$,
one finds from Eq.~(\ref{solutionev}) the single pole in $\vep$ 
in the logarithm of the jet function at two loops.
For the quark case this term is
\bea
- \frac{{\cal G}^{[q] (2)}_0}{8} 
+ \frac{\beta_0\, {\cal G}^{[q] (1)}{}'}{32}
&=& -\, \frac{3}{8}C_F^2\, 
\left[ \frac{1}{16} - \frac{1}{2}\zeta(2) +\zeta(3)\right]~~
\nonumber\\
&\ & \hspace{-14mm}- \frac{1}{16}C_AC_F \, \left [ \frac{961}{216} +
  \frac{11}{4}\zeta(2) - \frac{13}{2}\zeta(3)\right]
\nonumber\\
&\ & \hspace{-14mm}
+ \, \frac{1}{16}\, C_F T_F n_F\, \left[ \frac{65}{54} + \zeta(2)\right] \,,
\label{Enmvalues}
\eea
using values of ${\cal G}^{[q]}(\vep)$ from ref.~\cite{MVVquarkformfactor}.
Notice the contribution from the running
of the finite term at one loop, which appears as an 
${\cal O}(\vep)$ contribution in ${\cal G}^{[i] (1)}$.   

The one-loop soft anomalous dimension in color-generator form is
\bea
{\bf \Gamma}_{S_\f}^{(1)}
&=& 
\frac{1}{2}\sum_{i\in \f} \sum_{j\ne i}\, {\bf T}_i\cdot {\bf T}_j
  \,\ln\left(\frac{\mu^2}{-s_{ij}}\right) \,,
\label{GammaS1T}
\eea
where $s_{ij} =(p_i+p_j)^2$, with all momenta defined to flow into
(or out of) the amplitude.   The ${\bf T}_i$ 
are given explicitly by color generators in the representation
of parton $i$, multiplied by $\pm 1$:  plus one for an outgoing 
quark or gluon, or incoming antiquark;  minus one for an incoming 
quark or gluon, or outgoing antiquark.  
The color generator form for the anomalous dimension matrix
is more flexible, but less explicit, than the corresponding 
matrix expressions in a chosen basis of color tensors for the amplitude.  
An example of the latter for $q\bar{q}\to q\bar{q}$ scattering, 
in an $s$-channel $t$-channel singlet basis, is
\begin{equation}
\label{Eq:1l_gammaS}
{\bf \Gamma}_{S_\f}^{(1)}=
\left( \begin{array}{cc} \frac{1}{N_c}\left(\mathcal{U}-\mathcal{T}\right)
      +2\,C_F\,\mathcal{S} &
\left(\mathcal{S}-\mathcal{U}\right)\\ \\ 
\left(\mathcal{T}-\mathcal{U}\right) 
&\frac{1}{N_c}\left(\mathcal{U}-\mathcal{S}\right)+2\,C_F\,\mathcal{T}
\end{array} \right),
\end{equation}
where
$\mathcal{T}\equiv\ln\left(\frac{-t}{\mu^2}\right)$, and 
so on for the other Mandelstam invariants, defined by 
$s=s_{12}$, $t=s_{13}$, $u=s_{14}$.   
Resummed cross sections are determined by the eigenvalues 
and eigenvectors of these matrices~\cite{KOSjet,DM05}.  

\begin{figure}
\centerline{\epsfxsize=7cm \epsffile{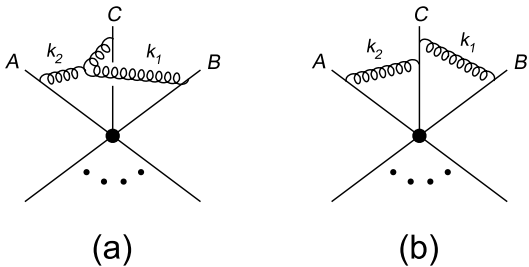}}
\caption{\label{3Efig}Two-loop diagrams  involving
three eikonal lines.}
\end{figure}

We are now ready to provide our result for the full two-loop soft 
anomalous dimension matrix,
\bea
{\bf \Gamma}_{S_\f}^{(2)} 
= \frac{K}{2}\, {\bf \Gamma}_{S_\f}^{(1)}\, .
\label{Gamma2}
\eea
Here $K=C_A(67/18 -\zeta(2)) - 10T_Fn_F/9$ is the same constant 
appearing in the relation between the one- and two-loop Sudakov,
or ``cusp'' anomalous dimensions~\cite{cusp,cross}: 
$\gamma_K^{[i]} = 2 C_i (\as/\pi) [ 1+(\as/\pi)K/2 ]$.
Remarkably, relationship~(\ref{Gamma2}) holds for an arbitrary
$2\to n$ process, even though the two-loop diagrams shown
in Fig.~\ref{3Efig} apparently couple together the color factors 
of three eikonal (Wilson) lines coherently.  
We derive Eq.~(\ref{Gamma2}) using the color generator formalism;
however, the result is completely general, and applies to explicit
matrix representations such as Eq.~(\ref{Eq:1l_gammaS}).

Following the method described in detail at one loop in Ref.~\cite{KOS},
and extended to two loops in Ref.~\cite{aybat06}, the two-loop 
anomalous dimension is found from the residue of single-pole terms 
in suitable combinations of Wilson lines computed at two loops.  
The simplicity of the result~(\ref{Gamma2}) follows from the 
special properties of the diagrams of Fig.~\ref{3Efig}, 
which connect three different eikonal lines.  
First consider Fig.~\ref{3Efig}a, where a three-gluon coupling ties together
three eikonals labelled $v_A$, $v_B$ and $v_C$,
in an otherwise arbitrary eikonal process.  We shall prove that this 
integral is zero, as long as we take the eikonals $v_A$ and $v_B$ 
to be lightlike.  
 
Since $v_A$ and $v_B$ are lightlike, we choose a frame in which 
$v_A^\mu=\delta^{\mu -}$ and $v_B^\mu = \delta^{\mu +}$.  Components 
transverse to $v_A$ and $v_B$ carry the subscript $T$.
The eikonal integral in momentum space is then
%
\begin{widetext}
\begin{eqnarray}
F_{1a}(v_A,v_B,v_C)
&=&
\int d^Dk_1 d^Dk_2\, 
\frac{1}{k_1^2+i\epsilon}\,\frac{1}{k_2^2+i\epsilon}
\frac{1}{(k_1+k_2)^2+i\epsilon}
\, \frac{1}{ k_1^-+i\epsilon}\,
\frac{1}{ k_2^++i\epsilon}
\nonumber\\
&& \hskip-20mm 
\times \,
\frac{\Big[ v_C^-  \left(k_1^+ - k_2^+\right)
+ v_C^+  \left(k_1^- - k_2^-\right)
- v_{C,T} \cdot \left(k_{1,T} - k_{2,T}\right)
+ v_C^+\,  \left(  k_1^- + 2k_2^- \right)
+ v_C^-\, \left( -2k_1^+ - k_2^+\right)\, \Big]}
{ v_C^-  \left(k_1^+ + k_2^+\right)
+ v_C^+  \left(k_1^- + k_2^-\right)
- v_{C,T} \cdot \left(k_{1,T}+k_{2,T} \right)
+ i\epsilon} 
\, ,
 \label{F1a}
\end{eqnarray}
\end{widetext}
where the term in square brackets is the three-gluon
vertex momentum factor.
We now introduce a change of variables (with unit Jacobean) 
from momenta $k_i^\mu$ to $\bar{k}_i^\mu$, 
%
\bea
\left (k_1^+,k_1^-,k_{1,T}\right)
&=& \left( \zeta \, \bar{k}_2^-\,,
        \, \zeta^{-1} \, \bar{k}_2^+\,,
           \, \bar{k}_{2,T}\right) \,,
\nonumber\\
\left (k_2^+,k_2^-,k_{2,T}\right)
&=& \left( \zeta \, \bar{k}_1^- \,,
        \, \zeta^{-1} \, \bar{k}_1^+ \,,
           \, \bar{k}_{1,T}\right)\,,
\label{changevars}
\eea
where $\zeta = v_C^+/v_C^-$.
It provides
an expression 
identical to $F_{1a}(v_A,v_B,v_C)$, Eq.~(\ref{F1a}),
but of the opposite sign.  The integral corresponding to 
Fig.~\ref{3Efig}a therefore vanishes.

Regarding Fig.~\ref{3Efig}b, the same change of variables yields 
$1/[(v_C\cdot (k_1+k_2))(v_C\cdot k_1)]=
1/[(v_C\cdot (\bar{k}_1+\bar{k}_2))(v_C\cdot \bar{k}_2)]$,
from which it is easy to show that this diagram reduces to
the product of one-loop diagrams, and so does not
contribute to the two-loop anomalous dimension.
Indeed, the only nontrivial contributions to ${\bf \Gamma}_{S_\f}^{(2)}$
at two loops involve only two eikonal lines.
Using results from refs.~\cite{cusp,cross},
the color structure of these contributions reduces to that 
of a single gluon exchange. The sum of the diagrams then modifies
the one-loop result by the same multiplicative factor
as for the cusp anomalous dimension, which gives Eq.~(\ref{Gamma2}).

The explicit expression for single poles in $2\to n$ amplitudes
is easily found from Eqs.~(\ref{solutionev}) and (\ref{expoS}) using
the explicit form of the two-loop matrix (\ref{Gamma2}),
\bea
\left | {\cal M}_\f^{(2)}\right\rangle^{\rm (single\ pole)}
&=& \label{Bornsingleexpand}\\
&\ & \hspace{-35mm}
\frac{1}{\vep}\, \left[
\sum_{i\in \f} \left( - \frac{{\cal G}^{[i](2)}_0}{8} 
+ \frac{\beta_0\, {\cal G}^{[i](1)}{}' }{32}\right)
+\frac{K}{8}\,
{\bf \Gamma}_{S_\f}^{(1)}\, \right] \, \left | {\cal M}_\f^{(0)}\right\rangle
\nonumber\\
&\ & \hspace{-35mm}
- \, \frac{1}{4\varepsilon}
 \sum_{i\in \f}{\cal G}_0^{[i](1)}\, \left | {\cal H}_\f^{(1)}(0)\right\rangle
- \sum_{i\in \f} \frac{1}{8\varepsilon}\, \gamma^{[i](1)}_K 
\, \left | {\cal H}_\f^{(1)}{}'(0)\right\rangle
\,.\nonumber
\eea
Here we normalize the one-loop hard scattering by
absorbing into it all finite terms from the jet functions, order by order,
and $|{\cal H}_\f^{(1)}(0)\rangle,\ |{\cal H}_\f^{(1)}{}'(0)\rangle$ are
this function and its derivative with respect to $\vep$,
respectively, evaluated at $\vep=0$.
(This absorption is possible to any loop order because the jets
are diagonal in color.)
Explicit comparison~\cite{aybat06} shows that this simple result
agrees with all single-pole terms found at $2\to 2$
in the literature, as summarized for
example in Refs.~\cite{BDD03,glover04}. 
It also predicts all such poles in a $2\to n$ process, once
the one-loop hard part $\left | {\cal H}_\f^{(1)}\right\rangle$
is known. 

We remark that an analogous anomalous dimension matrix for Wilson 
lines has been computed in Ref.~\cite{cross}, in the forward 
limit $t\to 0$.  This limit is a singular one, with respect to our 
arguments regarding Fig.~\ref{3Efig}; thus our results
and theirs are not directly comparable.

It is also worth remarking on the relationship between our
results and the influential alternative formalism of Ref.~\cite{catani98},
in which both pole and finite terms are put into an exponential
form to two loops.  We may think of these as alternative schemes
for organizing infrared poles.
When explicit calculations are organized according to the scheme
of Ref.~\cite{catani98}, more complex color products are found,
namely $i f_{abc}{\bf T}^a_i{\bf T}^b_j{\bf T}^c_k
 = - [ {\bf T}_i \cdot {\bf T}_j \, , \, {\bf T}_j \cdot {\bf T}_k ] $,
appearing in the matrix $\hat{\bf H}^{(2)}$ at order
$1/\vep$~\cite{BDD03,glover04}.
Such products are not encountered in the resummation scheme
described above.  
These differences in color structure, however, are by no 
means disagreements.  They arise from a particular commutator,
between the one-loop finite terms that are exponentiated
in the formalism of Ref.~\cite{catani98}, and the one-loop
soft anomalous dimension matrix ${\bf \Gamma}_{S_\f}^{(1)}$.
The result of performing the commutator~\cite{aybat06} 
agrees with the form of $\hat{\bf H}^{(2)}$ in Ref.~\cite{BDD03} 
for $2\to2$ processes, and with that proposed in Ref.~\cite{BDK04}
for 2~gluon~$\to~n$~gluon processes, based on consistency of
collinear limits.

Given an explicit two-loop amplitude, the strategy described 
here may be reversed, and ${\bf \Gamma}_{S_\f}^{(2)}$ 
extracted directly from the amplitude.  This approach was 
adopted in Ref.~\cite{jantzen05} for the case of quark-quark elastic scattering, in the context of electroweak Sudakov corrections.
The original version of Ref.\ \cite{jantzen05} differs from ours due to omission of the 
commutator contribution described above.  The authors have informed us that a revision 
is in preparation.

Similar remarks apply to the color-diagonal single poles
given in Eq.~(\ref{Bornsingleexpand}).  These coefficients
do not equal the corresponding coefficients $H_i^{(2)}$ 
in the formalism of Ref.~\cite{catani98}, but they are 
connected~\cite{RSvN04}.  The difference can be related precisely to the
different treatment of finite terms in the two approaches~\cite{aybat06}.

In addition to clarifying the structure of singular terms in 
calculations of $2\to 2$ processes at two loops, 
the results outlined here have potentially 
useful consequences and suggest further directions of research.  
Eq.~(\ref{Bornsingleexpand}) predicts the two-loop pole structure
for any $2\to n$ process, in color-generator form,
for any process whose one-loop hard function is known to ${\cal O}(\vep)$.

Another practical consequence
is that, because the one- and two-loop anomalous dimensions are
proportional, all terms in the expansion of the soft function commute
to next-to-next-to-leading level (NNLL), and in this approximation, the
ordering operator P can be dropped in Eq.~(\ref{expoS}).  
Thus, once the color eigenstates of the one-loop matrix are
known, the same states will diagonalize the two-loop matrix.
A semi-numerical approach, bypassing diagonalization, 
is to simply exponentiate the relevant matrices in any convenient
basis~\cite{BSZ}.  Given the relation~(\ref{Gamma2}), this is now 
possible at NNLL as well as NLL.

The study of these matrices for processes beyond $2\to 2$,
already begun in Ref.~\cite{kyrieleis05}, is clearly an important challenge.
Another intriguing question is whether the proportionality~(\ref{Gamma2})
might extend beyond two loops, whether in QCD or any of 
its allied gauge theories.
If so, it could have consequences for the interpretation
of infrared diverences in the relevant theory.  The extension, and/or
modification of the results above for the production of massive
colored particles is another important direction for research.

\acknowledgments
This work was supported in part
by the National Science Foundation, grants PHY-0098527 and PHY-0354776,
and by the Department of Energy under contract DE--AC02--76SF00515.
We thank the authors of Ref.\ \cite{jantzen05} for a very helpful exchange.
We also wish to thank Babis Anastasiou, Carola Berger, 
Zvi Bern, Yuri Dokshitzer, Nigel Glover, 
David Kosower, Gavin Salam, Jack Smith
and Werner Vogelsang for very helpful
conversations.   LD thanks the Kavli Institute for Theoretical
Physics and the Aspen Center for Physics for support during a portion of this work,
and GS thanks SLAC for hospitality.

\end{document}